\documentclass[twocolumn]{aastex631}

\usepackage{amsmath}
\usepackage{hyperref}
\usepackage[acronym]{glossaries}

\newacronym{gw}{GW}{gravitational wave}
\newacronym[plural=AGNs, firstplural=active galactic nuclei]{agn}{AGN}{active galactic nucleus}
\newacronym{sf}{SF}{structure function}

\definecolor{alizarin}{rgb}{0.82, 0.1, 0.26}
\definecolor{darkviolet}{rgb}{0.58, 0.0, 0.83}
\newcommand{\response}[1]{#1}

\begin{document}

\title{Multi-messenger constraints on LIGO/Virgo/KAGRA gravitational wave binary black holes merging in AGN disks}

\author[0000-0002-1270-7666]{T. Cabrera}
\affiliation{McWilliams Center for Cosmology and Astrophysics, Carnegie Mellon University,
5000 Forbes Avenue, Pittsburgh, PA 15213, USA}

\author[0000-0002-6011-0530]{A. Palmese}
\affiliation{McWilliams Center for Cosmology and Astrophysics, Carnegie Mellon University,
5000 Forbes Avenue, Pittsburgh, PA 15213, USA}

\author[0000-0002-1980-5293]{M. Fishbach}
\affiliation{Canadian Institute for Theoretical Astrophysics, 60 St George St, University of Toronto, Toronto, ON M5S 3H8, Canada}

\begin{abstract}
While the LIGO/Virgo/KAGRA (LVK) gravitational wave (GW) detectors have detected over 300 binary black hole (BBH) mergers to date, the first confirmation of an electromagnetic (EM) counterpart to such an event remains elusive.
Previous works have performed searches for counterpart candidates in transient catalogs and have identified active galactic nuclei (AGN) flares coincident with GW events; existing theory predicts that such flares may arise from the interaction of the merger remnant with the embedding accretion disk environment.
We apply a statistical formalism to measure the significance of coincidence for the catalog as a whole, measuring that less than 3\% (90\% credible interval) of LVK BBH mergers give rise to observable AGN flares.
This result still allows up to $\sim 40\%$ of BBH mergers to originate in AGN disks.
We also examine the individual coincidences of each merger/flare pairing, determining that in all cases the flares are more likely to belong to a background population of flares not associated with GW events.
Our results are consistent with theoretical predictions accounting for the observability of EM counterparts in AGN disks, as well as based on the fact that the most massive/luminous AGNs (such as those included in the search) are not expected to harbor the majority of the BBHs.
We emphasize that developing both the means to distinguish BBH counterpart flares from background AGN flares and an understanding of which BBHs are most likely to produce AGN flares as counterparts is critical to optimize the use of follow-up resources.
\end{abstract}

\section{Introduction}
\label{sec:intro}

Ninety gravitational wave (GW) events were reported in the LIGO/Virgo/KAGRA (LVK) GWTC-3 catalog covering the first three observing runs (O1-O3) of the detector network \citep{abbott_gwtc-1_2019,abbott_gwtc-2_2021,abbott_gwtc-3_2023,abbott_gwtc-3_2023-1}, and more than 200 additional candidate events have been detected during the ongoing fourth observing run (O4).
Over 95\% of these events are binary-black hole (BBH) mergers, and much effort has been dedicated to discerning the astrophysical origin of these objects (e.g. \citealt{raccanelli_determining_2016,eldridge_bpass_2016,hotokezaka_implications_2017,spera_merging_2019,marchant_pulsational_2019,di_carlo_binary_2020,rodriguez_observed_2021,mapelli_mass_2021,callister_state_2021,callister_who_2021,klencki_it_2021,godfrey_cosmic_2023,karathanasis_binary_2023}).
Significant discussion is focused on determining the contribution of various formation channels to the overall BBH population, with the current dataset supporting a combination of multiple formation channels (e.g. \citealt{arca_sedda_fingerprints_2020,Zevin2021,bouffanais_new_2021,mapelli_cosmic_2022,Cheng:2023,Ray:2024hos}). 

One such formation channel is the active galactic nucleus (AGN) channel.
In this channel, BBHs are formed from stars and BHs embedded in the accretion disks of AGNs.
The gaseous and dusty environment is expected to be one of the best accelerants for BBH formation and inspiral due to the dynamical friction manifested therein \citep{bartos_rapid_2017,stone_assisted_2017,tagawa_formation_2020,ishibashi_gravitational_2024,whitehead_gas_2024,wang_simulation_2025}.
Furthermore, the deep potential wells of the supermassive black holes (SMBHs) located at the centers of AGNs are conducive to retaining previous merger remnants and facilitate hierarchical mergers; such repeated mergers could contribute significantly to the observed high-mass ($>50~M_\sun$) and upper mass gap objects \citep{doctor_black_2020,tagawa_signatures_2021,ford_binary_2022,li_time-dependent_2022,vaccaro_impact_2024}, where a sub-population of mergers is observed \citep{MaganaHernandez:2024qkz,MaganaHernandez:2025cnu,antonini_star_2025, tong_evidence_2025}.

A special feature of the AGN formation channel is that it is thought to enable the production of electromagnetic (EM) counterparts following BBH mergers \citep{bartos_rapid_2017,perna_binary_2018,mckernan_ram-pressure_2019,perna_limits_2019,wang_accretion-modified_2021,tagawa_high-energy_2023,tagawa_shock_2024,ma_electromagnetic_2024,chen_electromagnetic_2024,rodriguez-ramirez_optical_2024}.
The detection of EM counterparts to GW events is the gold standard in the localization of these events; in addition, EM data greatly amplifies the  efficacy of analyses using GW data for measurements of key physical parameters such as the Hubble constant (see \citealt{GWCosmology} for a review).
To date, only the binary neutron star (BNS) merger GW170817 has been confidently associated with an EM counterpart \citep{abbottMultimessengerObservationsBinary2017}, and the latest estimations of BNS or neutron star-black hole multimessenger detection rates are converging towards a handful per year due to the low volumetric rates of these systems \citep{kunnumkai_GW230529,kunnumkai_O5}. On the other hand, despite a growing catalog of BBH alerts, no EM transient has been unequivocally connected to a BBH merger.
The most investigated BBH counterpart candidate is associated with the high-mass event GW190521 \citep{abbott_gw190521_2020}, and occurred roughly 50 days after the LVK trigger \citep{graham_candidate_2020}.

Leading models of BBH counterparts involve the interaction of the merger remnant with the AGN disk material, but the particularities of the manifestation of a counterpart are still uncertain.
Several models consider the formation of an accretion feedback-driven bubble about the remnant, which then breaks out from the surface of the accretion disk in a flare-like manner \citep{mckernan_ram-pressure_2019,wang_accretion-modified_2021,rodriguez-ramirez_optical_2024}; other works study more extreme possibilities such as that of an accretion-induced jet punching out of the disk \citep{tagawa_high-energy_2023,tagawa_shock_2024}.
This diversity of proposed counterpart mechanisms complicates the confirmation of such transients, as a wide variety of timescales and flare morphologies is possible.
Confirmation is further complicated by the necessary coincidence of these signals with the inherent variability of the host AGNs.
Although the understanding of AGN activity continues to advance, along with possible means to disentangle genuine transients from standard activity, the one-to-one matching of AGN flares to GW events has remained a milestone of future work.

Accepting the present uncertainties in these studies, several BBH counterpart candidates have been proposed by different teams  \citep{connaughton_fermi_2016,greiner_fermi-gbm_2016,connaughton_interpretation_2018,graham_candidate_2020,graham_light_2023,cabrera_searching_2024}.
One broad search was conducted by \citet{graham_light_2023}, which sifted data from the Zwicky Transient Facility (ZTF) \citep{bellm_zwicky_2019} to find AGN flares coincident with GW events from GWTC-3.
20 AGN flares with sufficiently interesting morphology were found among the catalog, 7 of which being coincident to a collective 9 GW events, to the extent that they could not be eliminated as potential counterparts to the respective events.

Although it is challenging to confidently associate AGN flares to individual BBH events, we can better understand the statistical association by considering the population of events.
While the AGN flares presented in \citet{graham_light_2023} have not been definitively identified as EM counterparts, this work undertakes the task of studying the statistical significance of such a dataset as it pertains to the association of LVK BHs with AGN flares.
\response{
Following \citealt{bartos_gravitational-wave_2017}, several previous works have investigated the link   between BBHs and AGN catalogs or AGN flares \citep{veronesi_most-luminous_2023, veronesi_constraining_2025, veronesi_agn-flares_2025, zhu_evidence_2025}.
In this work, we instead follow the methods presented in \citealt{palmese_ligovirgo_2021}, which adapted the neutrino-supernova association formalism of \citet{morgan_neutrino_2019} to the case of GW events and AGN flares; these methods enable measurement not only of the astrophysical fraction $\lambda$ of BBHs that produce AGN flares, but also association probabilities for each possible GW-AGN pairing, allowing for a consideration of which event pairings are most favored by spatiotemporal coincidence alone.
}
Here, we implement a quasar luminosity function (QLF) (from \citealt{hopkins_observational_2007}) and selection effects to better situate the analysis for observational data, and refine our background AGN flare rate calculations to consider differing AGN flare morphologies. In Section \ref{sec:methods} we describe the statistical method used and the assumptions made in the analysis. In Section \ref{sec:data} we describe the data used, while in Section \ref{sec:results} we show our results. Sections \ref{sec:discussion} and \ref{sec:conclusions} include a discussion of our findings and the conclusions from this work, respectively.
Throughout this paper we assume a cosmology where $H_0 = 70~{\rm km~s^{-1}~Mpc^{-1}}$ and $\Omega_m = 0.3$.

\section{Methods}\label{sec:methods}

The framework of \citet{palmese_ligovirgo_2021} considers the coincidence of an observed set of AGN flares in the context of a GW event catalog and an AGN flare background model to investigate the scenario where some flares originate from BBH mergers.
In other words, if a fraction of the population of AGN flares observed in coincidence with GW events is indeed related to the compact object mergers, they will follow a different distribution in sky position and redshift compared to the population of AGN flares arising from unrelated phenomena.
We can statistically constrain that fraction through the method described in \citet{palmese_ligovirgo_2021}, and we summarize the main components of the inference in what follows.

For GW event $i$, the likelihood of observing $k$ AGN flares at sky positions and redshifts $\{ \Omega^{\rm AGN}_{ij}, z^{\rm AGN}_{ij} \}^{k}_{j=1}$ and the GW data $x^{\rm GW}_i$, given the astrophysical fraction of LVK-observable BBHs that produce AGN flares $\lambda$ and the AGN flare background rate $R_B$, is:
\begin{widetext}
\begin{equation}
\begin{split}
    \mathcal{L}_i
        &\equiv
        p \left(
            \left\{
                \Omega_{ij}^{\rm AGN},
                z_{ij}^{\rm AGN}
            \right\}_{j=1}^{k},
            x_i^{\rm GW}
            |
            \lambda,
            R_B
        \right) \\
    &= \int d\Omega_i^{\rm GW} dz_i^{\rm GW}
        p \left(
            \left\{
                \Omega_{ij}^{\rm AGN},
                z_{ij}^{\rm AGN}
            \right\}_{j=1}^{k},
            \Omega_i^{\rm GW},
            z_i^{\rm GW},
            x_i^{\rm GW}
            |
            \lambda,
            R_B
        \right)
\end{split}\label{eq:Li}
\end{equation}
\end{widetext}
where in the second line we have marginalized over the uncertain GW sky position $\Omega_i^{\rm GW}$ and redshift $z_i^{\rm GW}$.

Assuming that the AGN flares are generated by independent processes,
the integrand of the equation above can be expressed using the formalism of an inhomogeneous Poisson process similarly to \citet{mandel_extracting_2019}:
\begin{widetext}
\begin{equation}
\begin{split}
    & p \left(
        \left\{
            \Omega_{ij}^{\rm AGN},
            z_{ij}^{\rm AGN}
        \right\}_{j=1}^{k},
        \Omega_i^{\rm GW},
        z_i^{\rm GW},
        x_i^{\rm GW}
        |
        \lambda,
        R_B
    \right) \\
    &= \prod_{j=1}^{k}
        p \left(
            x_i^{\rm GW}
            |
            \Omega_i^{\rm GW},
            z_i^{\rm GW}
        \right)
        p_0 \left(
            \Omega_i^{\rm GW},
            z_i^{\rm GW}
        \right)
        \frac{dN}{d\Omega dz} \left(
            \Omega_{ij}^{\rm AGN},
            z_{ij}^{\rm AGN}
            |
            \Omega_i^{\rm GW},
            z_i^{\rm GW},
            \lambda,
            R_B
        \right)
    e^{-\mu_i}
    \label{eq:poisson}
\end{split}
\end{equation}
\end{widetext}
where $p_0 (\Omega, z)$ is the prior on the GW event position and $dN / (d\Omega dz)$ is the distribution of AGN flares.
\response{In Eq. (\ref{eq:poisson}), $dN / (d\Omega dz)$ is evaluated at the position of the $j^{\rm th}$ AGN flare ($\Omega_{ij}^{\rm AGN}, z_{ij}^{\rm AGN}$), and depends on the (unknown) position of the GW ($\Omega_{i}^{\rm GW}, z_{i}^{\rm GW}$), the BBH flare production fraction $\lambda$, and the background distribution $R_B$ (the latter in units of flares per steradian per redshift per follow-up time window):}
\begin{widetext}
\begin{equation}
    \frac{dN}{d\Omega dz} \left(
        \Omega,
        z
        |
        \Omega_i^{\rm GW},
        z_i^{\rm GW},
        \lambda,
        R_B
    \right)
    =
    \lambda \delta\left( \Omega_i^{\rm GW} - \Omega \right) \delta\left( z_i^{\rm GW} - z \right)
    + R_B(\Omega, z),
\label{eq:dist}
\end{equation}
\end{widetext}
where the $\delta$ represent Dirac $\delta$ functions. 
In Eq. (\ref{eq:poisson}), $\mu_i$ is the expected number of flares associated with the GW event, taking into account the detection probability $P^{\rm AGN}_{\rm det}$:
\begin{equation}
    \mu_i
    \equiv
    \int d\Omega dz
    P^{\rm AGN}_{\rm det}(\Omega, z)
    \frac{dN}{d\Omega dz}.
\label{eq:mu}
\end{equation}
After performing the integration, the likelihood in Eq. (\ref{eq:Li}) for a single GW event takes the form
\begin{widetext}
\begin{equation}
\mathcal{L}_i 
\propto
\prod_{j = 1}^{k} \left[
    \lambda p(\Omega_{ij}^\mathrm{AGN}, z_{ij}^\mathrm{AGN} \mid x_i^\mathrm{GW})
    + R_B \left( \Omega_{j}^\mathrm{AGN}, z_{j}^\mathrm{AGN} \right)
\right] e^{-\mu_i}.
\label{eq:one_event_likelihood}
\end{equation}
\end{widetext}

\response{
A posterior on the desired parameters (in this case $\lambda$) can be constructed from the likelihoods for all GW events considered during a follow-up campaign by multiplying the product of the single-event likelihoods by a suitable prior on $\lambda$ (in this work we use a uniform distribution over $\lambda \in [0, 1]$).
The combination of the single-event likelihoods in this way is permitted by the fact that the selections of coincident AGN flares for all GW events are independent of each other: while the follow-up window for a particular GW event may contain genuine counterpart flares to other GW events, the inclusion of those flares in the list of coincidences is not dependent on the detection of the other GW events (in fact, some included flares may be counterparts to BBH mergers not detected by LVK). Note that while there are differences, this aspect has a similar application in the dark standard siren galaxy catalog approach \citep{schutz1986determining,chenTwoCentHubble2018,palmese_standard_2023}. Also in that case, the galaxy catalog is the same for all GW events, and the GW likelihoods are multiplied. The formalism only becomes more complicated when the galaxies' redshifts are imperfectly known (e.g. for photometric redshifts, which we do not consider here), but the GW likelihoods may still be multiplied \citep{Hitchhicker_guide}.
The final formulation of the posterior on $\lambda$ takes the form:}
\begin{equation}
\begin{split}
    p & \left(
        \lambda
        \mid
        \{
            x_i^{\rm GW}
        \}_{i=1}^N,
        \{
            \{
                \Omega_{ij}^{\rm AGN},
                z_{ij}^{\rm AGN}
            \}_{j=1}^{k}
        \}_{i=1}^{N},
        R_B
    \right) \\ 
    & \propto p(\lambda) \prod_{i=1}^N \mathcal{L}_i.
\label{eq:posterior}
\end{split}
\end{equation}
\response{
The two driving components of this analysis are the GW posterior probabilities at the locations of the AGN flares $p(\Omega_{ij}^\mathrm{AGN}, z_{ij}^\mathrm{AGN} \mid x_i^\mathrm{GW})$ and the background flare rate $R_B$ at the same locations.
The relative magnitudes of the components determines whether it is more likely for a flare to have come from the GW event in question or a background distribution of flares, and accordingly determine the shape of the posterior in the case of coincident flares.
Note that for GW events with no coincident flares, the likelihood reduces to the term associated with the background rate, which simplifies to
\begin{equation}
    \label{eq:flareless}
    \mathcal{L}_{i,{\rm flareless}} \propto e^{-\mu_i}.
\end{equation}
}

\section{Data}\label{sec:data}

\subsection{GW event data}

\response{
We use GWTC-3.0 \citep{abbott_gwtc-3_2023,abbott_gwtc-3_2023-1} to assemble our set of GW events for this analysis (events from LVK O1 and O2 do not have ZTF coverage).
Of the 83 compact binary coalescences in this catalog, 5 (GW190425, GW190426\_152155, GW191219\_163120, GW200105\_162426, and GW200115\_042309) have significant probability of involving at least one NS, and 2 (GW190424\_180648 and GW190909\_114149) are reported with lowered significance in the most recent version of the catalog.
Removing these 7 events leaves us with 76 BBH mergers, which we use as our GW sample for this analysis; these are listed in Table \ref{tab:gws}.
We use IMRPhenomXPHM waveform skymaps for all events.
}

\subsection{AGN flare data}

\response{
\citealt{graham_light_2023} identified 20 AGN flares in their ZTF dataset, including cuts on flare morphology, energetics, and color evolution; these cuts served to identify a set of AGN flares with ambiguous origin, i.e. those not clearly indicative of a tidal disruption event or a supernova.
Of these 20 flares, two (J150748.68+723506.1 and J234420.76+471828.9) do not have published redshifts, and therefore a 3D crossmatch to GW skymaps is not possible.
Additionally, the host of flare J053408.42+085450.7 has exhibited blazar-like activity in long-term ZTF monitoring since 2023, and so there is a significant chance that this flare did not originate as a BBH counterpart.
We exclude these three flares, leaving 17 flares for use in our analysis.
}

\response{
We determine flare onset times by fitting to the lightcurve of each flare in flux.
Using ZTF lightcurves \citep{masci_zwicky_2019} accessed through the NASA/IPAC Infrared Science Archive\footnote{\url{https://irsa.ipac.caltech.edu/Missions/ztf.html}}, we fit a Gaussian rise-exponential decay model to the locale of each flare; the model takes the form
\begin{equation}
    f(t) = \begin{cases}
        f_0 + f_{\rm peak} e^{- \frac{1}{2} \left(\frac{t - {\rm MJD}_{\rm peak}}{\sigma_{\rm rise}}\right)^2}, & t < {\rm MJD}_{\rm peak} \\
        f_0 + f_{\rm peak} e^{-\frac{t - {\rm MJD}_{\rm peak}}{\sigma_{\rm decay}}}, & t \geq {\rm MJD}_{\rm peak}
    \end{cases}.
\end{equation}
We calculate the flare onset time ${\rm MJD}_0$ as a function of the peak flare time ${\rm MJD}_{\rm peak}$ and the rise timescale $\sigma_{\rm rise}$:
\begin{equation}
    {\rm MJD}_0 = {\rm MJD}_{\rm peak} - 3 \sigma_{\rm rise}.
\end{equation}
These three parameters are presented for each flare in Table \ref{tab:flares}, along with the redshift of each flare.
}

\subsection{Associating GWs and AGN flares}

\response{
We associate AGN flares with GW events by combining a spatial and a temporal crossmatch of the respective events.
The 3D spatial crossmatch is performed using a custom version of the \texttt{crossmatch} function of the \texttt{ligo.skymap.postprocess.crossmatch} module \citep{singerGoingDistanceMapping2016,singer_supplement_2016} which has been updated to allow for variable cosmology (we explored the effect of different cosmological parameters on the results of this work, and found an insignificant dependence on $H_0$ and $\Omega_m$ over flat $\Lambda$CDM models near our assumed cosmology, including the values from \emph{Planck} and SH0ES; \citealt{Planck:2018vyg,Riess:2021jrx}).
The temporal crossmatch is performed by comparing the flare onset time ${\rm MJD}_0$ to the GW trigger time.
A flare is considered coincident with a GW event if
\begin{enumerate}
    \item the flare is within the 90\% CI volume of the event, and
    \item the flare onset time is after and within 200 days of the GW trigger time.
\end{enumerate}
}

\begin{figure*}
    \centering
    \includegraphics[width=0.95\linewidth]{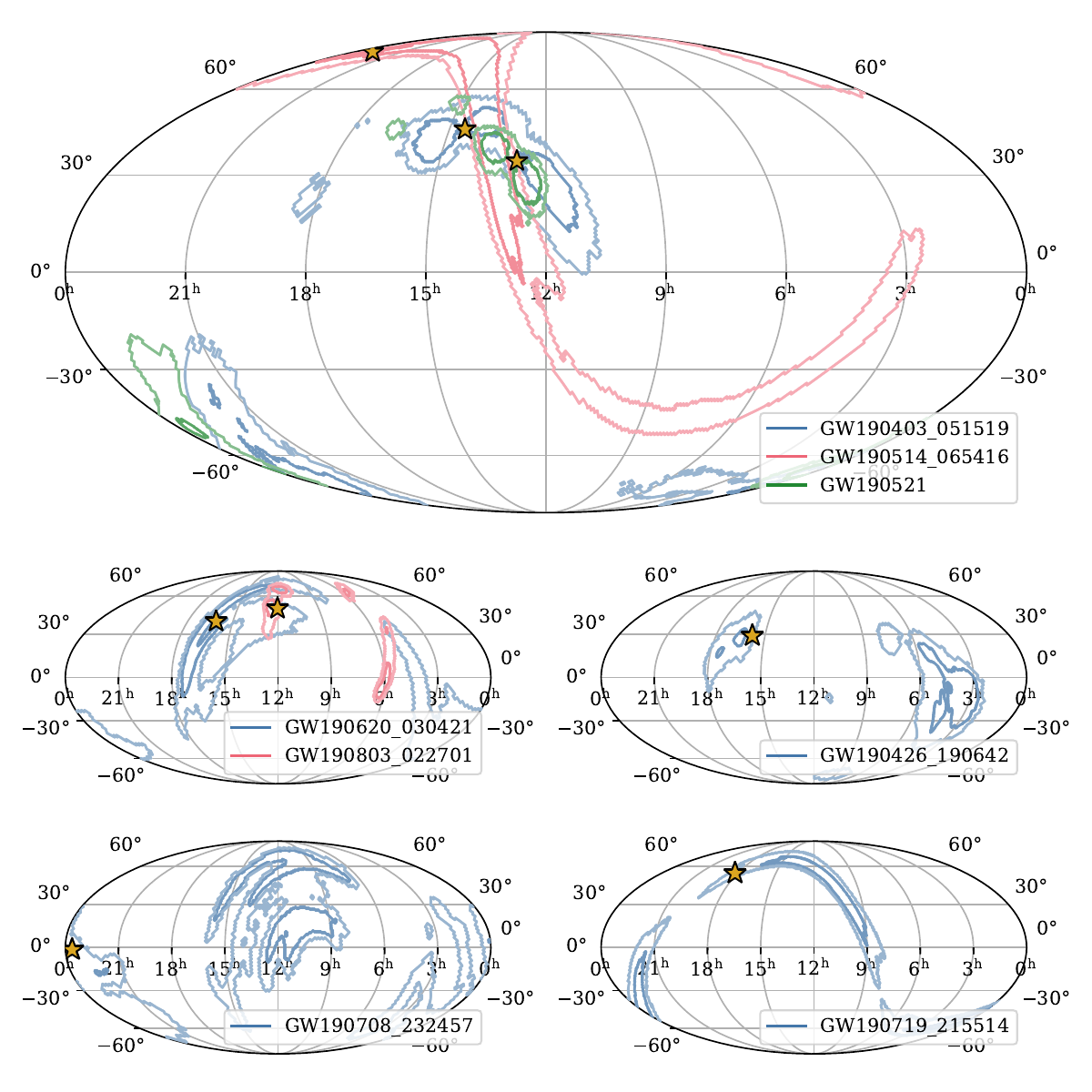}
    \caption{
        \response{
        Skymaps for the GW events coincident with at least one AGN flare.
        Each skymap plot includes the 2D 50\% and 90\% contours (with with the 50\% contour being the darker of the two).
        Coincident flares are shown as stars.
        GWs that share a coincident flare are shown in the same plot; see Figure \ref{fig:assoc_probs} for the exact GW-flare pairings.
        }
    }
    \label{fig:skymaps}
\end{figure*}

\response{
This procedure identifies 11 GW-flare associations in our dataset, which contain 8 unique GW events and 8 unique AGN flares; when evaluating our posterior, the sum over $k$ (sum over flares) in Eq. \ref{eq:one_event_likelihood} includes only those flares associated with the respective GW event.
The skymaps for the 8 GW events with coincident flares are plotted in Figure \ref{fig:skymaps}, along with the respectively associated flares.
}

\subsection{AGN flare background distributions}
\label{subsec:agnbkg}

\response{
We deconstruct the background distribution of AGN flares $R_B$ as a product of the physical distribution of AGNs, the rate of flares per AGN, and the follow-up time interval; assuming the distribution is isotropic over the sky, $R_B$ may be written as a function of redshift:
\begin{equation}
    R_B (z)
    =
    \frac{1}{4\pi}
    \cdot
    \frac{dN_{\rm AGN}}{dz}(z)
    \cdot
    r_{\rm flare}
    \cdot
    200~{\rm days}.
\end{equation}
Here, $dN_{\rm AGN}/dz$ is the distribution of AGN as a function of redshift, $r_{\rm flare}$ the rate of flares per AGN per unit time, and 200 days is the GW follow-up window used by \citealt{graham_light_2023} to find coincident AGN flares.
}

\response{
We use two different AGN distributions in different parts of our analysis.
As $dN/(d\Omega dz)$ is defined as the \textit{astrophysical} distribution of AGN flares (Eq. \ref{eq:dist}), the value of $R_B$ used here must be similarly astrophysical.
We define our fiducial distribution by applying a by applying a $<$20.5 $g$-band magnitude cut (the approximate depth of the ZTF survey) to the quasar luminosity function of \citep{hopkins_observational_2007}.
We also investigated other alternatives including a constant comoving AGN density of $10^{-4.75}$ AGN/Mpc$^3$ \citep{greene_mass_2007,greene_erratum_2009}, and a cut on bolometric luminosity informed by simulations of BBH mergers in AGN disks; the latter alternative is detailed in \S\ref{subsec:subsets}, and neither variation yielded a significant change in results.
}

\response{
In the calculation of the expected number of coincident flares $\mu_i$ (Eq. \ref{eq:mu}), the distribution of flares is multiplied by the detection probability $P_{\rm det}^{\rm AGN}$, producing a term representing the distribution of \textit{observed} AGN flares.
As we define $\lambda$ as the fraction of LVK BBHs that produce \textit{observable} flares, $P_{\rm det}^{\rm AGN}$ does not depend on the flare morphology; however, because the flare sample is constructed by crossmatching to an AGN catalog, the detection probability becomes linked to the completeness of the catalog.
Therefore, the combined term $P_{\rm det}^{\rm AGN} \cdot dN_{\rm AGN}/(d\Omega dz)$ can be taken as the distribution of AGN from the catalog of use.
As the \citealt{graham_light_2023} flare sample is constructed via crossmatch with the Million Quasars Catalog v7.2 \citep{flesch_million_2021}, we use the distribution of AGN for the catalog for $P_{\rm det}^{\rm AGN} \cdot dN_{\rm AGN}/(d\Omega dz)$; the integrand $P_{\rm det}^{\rm AGN} \cdot dN/(d\Omega dz)$ in Eq. \ref{eq:mu} can therefore be calculated by multiplying this distribution by the rate of flares per AGN per follow-up time window.
}

\response{
The flare rate $r_{\rm flare}$ is taken from \citet{graham_light_2023} as an empirical rate of $1.06 \pm 3.73 \times 10^{-8}$ flares per AGN per day.
This rate is constructed from the fraction of flares passing their morphological selection cuts, and is interpreted as the rate at which AGNs produce flares of this quality, regardless of the actual generating mechanism.
\response{Accordingly, this rate carries information on the generic background rate of counterpart-like AGN flares, not just those coincident with LVK events.}
We implement this flare rate in our analysis by sampling a rate from a normal distribution with a mean and standard deviation equal to the rate and uncertainty given by \citet{graham_light_2023}, truncated at $r_{\rm flare} = 0$ to ensure a non-negative rate.
We marginalize over this parameter in our final results.
}

\section{Results}\label{sec:results}

We evaluate our posterior by performing MCMC sampling over the $\lambda$ parameter with the \texttt{emcee} Python package \citep{foreman-mackey_emcee_2013}.
The code used for this is available on GitHub\footnote{\url{https://github.com/tomas-cabrera/bbhagn}}, along with the code used to produce all figures and tables.

\subsection{Fiducial analysis}

\begin{figure}
    \centering
    \includegraphics[width=1.0\linewidth]{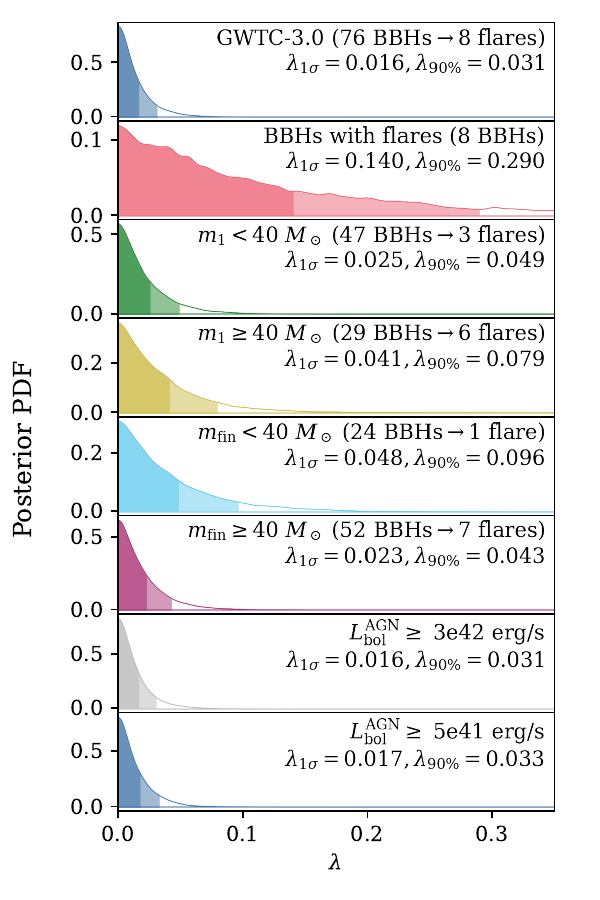}
    \caption{
        The posterior distributions on the flare production fraction $\lambda$ for LVK BBH mergers.
        The 1$\sigma$ and 90\% upper limit regions are shaded, with the respective thresholds labeled.
        The first posterior (dark blue) is the fiducial analysis, utilizing all 76 confident BBHs from GWTC-3.0 and applying an $m_g < 20.5$ cut to select AGN observable by ZTF.
        The second posterior (pink) uses only the 8 BBHs with coincident AGN flares, excluding the information provided by BBHs with no candidate counterparts.
        The third through sixth posteriors (green, yellow, light blue, purple) use BBHs passing the specified mass cuts; the number of BBHs used is included in plot label, as well as the number of flares associated with at least one BBH in the subset.
        The seventh and eighth posteriors (gray and dark blue) include all BBHs, but select AGN by bolometric luminosity $L_{\rm bol}^{\rm AGN}$ cuts instead of apparent magnitude (see \S\ref{subsec:subsets} for details).
        \response{
        The distributions are not plotted with the same vertical axis limits in order that the shapes of the distributions remain perceptible; the vertical axis range is labeled to allow for the comparison of subplot scales.
        }
    }
    \label{fig:lambda_posteriors}
\end{figure}

The resulting posteriors on $\lambda$ are displayed in Figure \ref{fig:lambda_posteriors}; in our fiducial analysis we find that the posterior probability is maximized for $\lambda = 0$, with a 90\% highest-probability upper limit of $\lambda < 3.1\%$.

We additionally combine the terms from Eq. \ref{eq:one_event_likelihood} to produce association probabilities describing the relative probability that a flare is caused by a particular BBH merger.
For a GW event $i$ and an AGN flare $j$, we calculate the association probability  as
\begin{widetext}
\begin{equation}
p^{\rm GW-AGN}_{ij}
=
\frac{
    \lambda p(\Omega_{ij}^\mathrm{AGN}, z_{ij}^\mathrm{AGN} \mid d_i^\mathrm{GW})
}{
    \sum_{i} \left( \lambda p(\Omega_{ij}^\mathrm{AGN}, z_{ij}^\mathrm{AGN} \mid d_i^\mathrm{GW}) \right)
    + R_B \left(\Omega_{j}^\mathrm{AGN}, z_{j}^\mathrm{AGN} \right)
},
\label{eq:prob}
\end{equation}
\end{widetext}
where $p(\Omega_{ij}^\mathrm{AGN}, z_{ij}^\mathrm{AGN} \mid d_i^\mathrm{GW})$ and $R_B$ are calculated as in Eq. \ref{eq:one_event_likelihood}.
This statistic weighs the flare $j$ - GW event $i$ association against all other possible sources for that flare, including any other coincident GW events and the background flare distribution.
We also calculate
\begin{widetext}
\begin{equation}
p^{\rm BG-AGN}_{j}
=
\frac{
    R_B \left(\Omega_{j}^\mathrm{AGN}, z_{j}^\mathrm{AGN} \right)
}{
    \sum_{i} \left( \lambda p(\Omega_{ij}^\mathrm{AGN}, z_{ij}^\mathrm{AGN} \mid d_i^\mathrm{GW}) \right)
    + R_B \left(\Omega_{j}^\mathrm{AGN}, z_{j}^\mathrm{AGN} \right)
},
\label{eq:probbg}
\end{equation}
\end{widetext}
which represents the relative probability that the flare $j$ came from the AGN background and not from any coincident GW event.

\begin{figure*}
    \centering
    \includegraphics[width=1.0\linewidth]{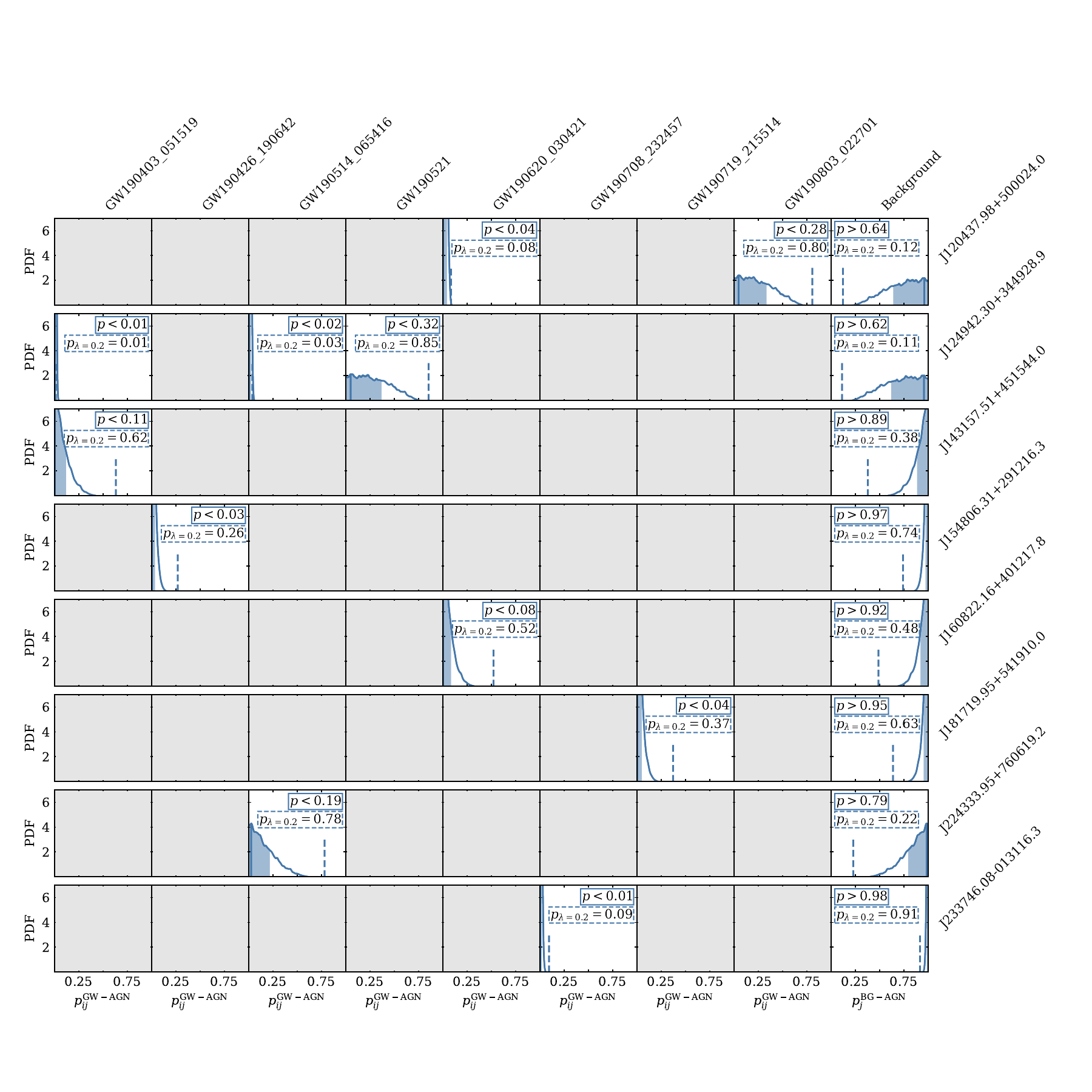}
    \caption{
        Posterior probabilities of association for individual GW-flare pairings.
        The subplots for the associations found in this study show the respective posterior PDFs (Eq. \ref{eq:prob}); all other subplots are grayed out.
        The rightmost column plots a comparable statistic showing the probability the flare came from a background AGN and not any of the coincident GW events (Eq. \ref{eq:probbg}).
        The 1$\sigma$ regions are shaded, with the values printed and outlined in a solid box (generally these are upper limits).
        The respective probabilities in the $\lambda = 0.2$ case are plotted as dashed lines, with the respective values printed above.
        Note that because the same $\lambda$ samples are used for all plots, the distributions have similar fine structure.
    }
    \label{fig:assoc_probs}
\end{figure*}

The posterior PDFs on these association probabilities are plotted in Fig. \ref{fig:assoc_probs}, in a grid mapping the possible GW event-AGN flare pairings, with an additional rightmost column plotting the background origin posteriors.
We calculate $p^{\rm GW-AGN}_{ij}$ and $p^{\rm BG-AGN}_{j}$ in two cases: one where we consider our inferred $\lambda$ posterior from our analysis including all 76 BBHs (substituting it for $\lambda$ in Eqs.~\ref{eq:prob} and \ref{eq:probbg}), and one where we set $\lambda = 0.2$ (vertical dashed lines). The $\lambda = 0.2$ value is a choice motivated by the findings of theoretical predictions: \citet{ford_binary_2022} find that 25-80\% of LVK BBH mergers could be originating in the disks of AGNs, while \citet{gayathri_gravitational_2023} find that the fraction should be about 20\%.
Even if they originate in AGN disks, we only expect a fraction of those events to give rise to observable EM counterparts, since the remnant may be kicked away from the observer, a jet may not form, the jet may be pointed away from the observer, or the flare may be obscured by the AGN dust; the fraction of the BBH mergers in AGN disks that gives rise to EM emission will depend on the still uncertain EM radiation mechanism and the observing geometry of the host.
Therefore, $20\%$ should be regarded as an upper limit for our value of $\lambda$.

\subsection{Auxiliary analyses}\label{subsec:subsets}

We also calculate the $\lambda$ posterior for sub-populations of our sample; these posteriors are also plotted in Figure \ref{fig:lambda_posteriors}.
Our first modified inference considers only the BBHs coincident with flares (8 BBHs); in this case, $\lambda$ is interpreted as the fraction of BBHs with coincident flares that actually produced a counterpart flare.
This analysis yields an upper limit of $\lambda < 29\%$, indicating that up to this fraction of BBHs with coincident AGN flares are genuine multimessenger sources.

We next introduce BBH mass cuts to examine whether $\lambda$ has any dependence on BBH mass.
We run our analysis using separately ``low"- and ``high"-mass BBHs, specifically using BBHs with median primary component masses $m_1$ less/greater than 40 $M_\odot$, and median final masses $m_{\rm fin}$ less/greater than the same threshold.
The primary mass cut is motivated by the understanding that black holes in the upper mass gap are more likely to originate from hierarchical mergers than isolated binaries, with the pair-instability supernova mass gap recently identified to be within $40-50~M_\odot$ \citep{MaganaHernandez:2025fkm,antonini2025gravitationalwavesrevealpairinstability,tong_evidence_2025}, and predictions indeed showing that the AGN formation channel is expected to significantly contribute to the merger rate of binaries with $m_1>40~M_\odot$ \citep{gayathri_black_2021}.
The final mass cut is independently motivated by identifying an appropriate threshold with which to split the bimodal distribution of final masses from GWTC-3.0.
\response{
In this study, both of these mass cuts result in the identification of high-mass BBH populations with more associated flares and GW-flare coincidences than their low-mass counterparts: the high primary mass population is has a total of 8 coincidences with 6 of the AGN flares versus 3 coincidences with 3 flares in the low primary mass population, and the high final mass population has 10 coincidences with 7 flares versus 1 coincidence with 1 flare in the low final mass population.
While these differences in high/low-mass associations would be considerably interesting if evident of an increased flare production rate from high-mass BBHs, the present statistics are insensitive to the tendency of high-mass BBHs to have larger localization volumes: higher mass BBHs can be detected at larger distances than low-mass events, and the subsequent larger volume leads to an increased rate of chance coincidence.
}

To investigate the dependence of $\lambda$ on AGN luminosity, we first take the AGN mass predicted to produce the most BBH mergers from \citealt{rowan_black_2024}, specifically $M_{\rm SMBH} = 10^7~ M_\odot$.
We then convert this mass into luminosity via the AGN mass-luminosity relationships in \citealt{wandel_central_1999} and \citealt{kaspi_reverberation_2000}, finding thresholds of $L^{\rm AGN}_{\rm bol} \sim 3 \cdot 10^42$ and $\sim 5 \cdot 10^41$ erg s$^{-1}$, respectively.
We then recalculate the $\lambda$ posterior in both cases, including AGN brighter than the respective bolometric luminosity value instead of the ZTF-motiviated magnitude cut for our fiducial study.
While these cuts cause significant changes in background AGN density ($\sim$1-2 orders of magnitude), the posteriors for these analyses are largely the same as that of our first fiducial analysis.

\section{Discussion}\label{sec:discussion}

Our posterior on $\lambda$ is consistent with the $\lambda = 0$ case, and leads to the expectation that $<3\%$ of LVK BBHs produce detectable AGN flares as counterparts, given the ZTF observations and the GWTC-3 catalog.
This is not surprising in light of the findings of previous works \citealt{graham_light_2023}: assuming that $\sim 50\%$ of LVK BBH mergers arise in AGN disks, that roughly half of the merger remnants are kicked away from the observer (therefore any EM radiation would be unlikely to reach us through the optically thick disk), and that about half of the AGNs are unobscured, they calculated an expected number of flares from BBH mergers of $\sim 3$, corresponding to less than $\sim 4\%$ of the GWTC-3 events under consideration.
With a similar reasoning, we can see that even if 40\% of LVK BBH mergers arise in AGN disks, we can get a value of $\lambda$ consistent with our upper limit.
\response{
In other words, our constraint is consistent with the scenario where $\lesssim40\%$ of LVK BBH events originate in the disks of AGNs, with geometrical constrains on flare visibility reducing the rate to the value we measure for our upper limit.
}

The association posterior probabilities in Fig. \ref{fig:assoc_probs} aid in the evaluation of causality in the case of particular multimessenger GW-flare pairings.
\response{
At the broadest level, it can be seen that the background case is preferred for all flares, exhibited by the $p_{j}^{\rm BG-AGN}$ distribution peaking at 1 and all of the $p_{ij}^{\rm GW-AGN}$ distributions peaking at 0.
}

\response{
In pursuit of causal connections, the most likely multimessenger pairings are GW190521 - J124942.30+344928.9 and GW190803\_022701 - J120437.98+500024.0 ($p_{ij}^{\rm GW-AGN} <$ 0.32 and 0.28, respectively).
The first association has been subject to continued discussion since its discovery \citep{graham_candidate_2020}, with the high mass and negative $\chi_{\rm eff}$ of the BBH adding additional evidence in favor of the dynamic, hierarchical, and AGN origins \citep{fragione_on_2020, liu_hierarchical_2021, li_multimessenger_2025}.
While less massive and with a more agnostic $\chi_{\rm eff}$, GW190803 nonetheless has favorable properites for the AGN origin: the component BHs are of masses appropriate for compact objects merging in AGN disks, which are typically expected to contribute to the $\sim 30~M_\odot$ feature in the mass distribution of BBHs and above more than to the $\sim 10~M_\odot$ feature \citep{gayathri_black_2021,mckernan_mcfacts_2024,Rowan:2024lla}, and the ambiguous $\chi_{\rm eff}$ still lacks the preference for positive values evident of isolated BBH formation.
It must be noted that in both cases the consideration of the $\chi_{\rm eff}$ parameter as indicative of a dynamic formation history is complicated in the in the case of AGN: while the AGN disk is expected to accelerate embedded BBH mergers to timescales shorter than that for isolated BBH spin alignment \citep{McKernan_2022}, there are other works that predict that the same environment can also accelerate spin alignment \citep{tagawa_spin_2020,li_spin_2022,vaccaro_impact_2024}.
It is similarly important to note that our analysis still favors the background flare origin for these flares ($p_{i}^{\rm BG-AGN} \gtrsim 0.6$), and that previous work finds insufficient evidence for confident association in the case of one of our pairings \citep{ashton_current_2021}.
}

\response{
The preference for non-association also receives some support from our auxiliary analyses, foremost because none of these inferences yield a lower limit on $\lambda$; the respective upper limits can be interpreted as defining the maximum fraction of each population that produce observable AGN flares.
While these inferences find greater upper limits than our fiducial analysis, we understand this behavior as arising solely from the number of BBHs used in each inference, and not indicative that any of the subpopulations are more likely to produce AGN flares.
We believe this because 1) complementary subpopulations (e.g. $m_1 < 40 M_\odot$ and $m_1 \ge 40 M_\odot$) both yield higher upper limits than the full population, instead of upper limits that can be averaged to the limit derived with the full population, and 2) across the different inferences, $\lambda_{\rm upper} \sim 1 / n_{\rm BBH}$.
We expect that an expanded dataset, in which both the full population and all subpopulations contain appreciably larger numbers of events, will resolve this present limitation on measuring $\lambda$.
}


\section{Conclusions}\label{sec:conclusions}

In this work we apply the BBH-AGN association formalism of \citet{palmese_ligovirgo_2021} and apply it to the AGN flare GW counterpart candidates of \citet{graham_light_2023}.
In doing so we measure the fraction of observed LVK GW events that produce detectable AGN flares $\lambda$, finding that less than 3\% of events produce flares with 90\% confidence. This result still leaves open the possibility that $\lesssim40\%$ of LVK BBH mergers originate in AGN disks. 
In addition we examine the particular pairings of GW events to AGN flares, and identify the most likely pairings as GW190521 - J124942.30+344928.9 and GW190803\_022701 - J120437.98+500024.0, albeit the probabilities of association remain $\sim$30\%.
We also note that these GW event has component masses favorable for the AGN channel, and $\chi_{\rm eff}$s that may indicate a dynamical formation history.

Altogether, our results indicate that multimessenger LVK BBH mergers with AGN flares as counterparts are rare, with likely less than one merger in $\sim$40 producing a counterpart of this type.
Our findings also indicate that the majority of candidate counterparts to BBH events are actually due to AGN activity unrelated to the GW event (e.g. explosive events breaking out from within the disk or disk instabilities close to the SMBH innermost stable circular orbit; \citealt{graham_light_2023}), supporting the need for the inclusion of a background population of AGN flares when associating them to BBH mergers, as well as motivating follow up observations of those to clarify their origin and better characterize their occurrence rates.
Moreover, we find no compelling evidence for high mass mergers to be more likely to be associated with the available AGN flares, although this is likely still limited by the present dataset.

Some remaining caveats exist for this measurement, primarily concerning the completeness of the AGN flare sample:
First, since models for BBH counterparts have not converged on a consistent phenomenology, it is unknown whether the ZTF data is deep enough to detect all GW-originated flares.
In addition, the AGN catalogs used to identify AGN-hosted flares have some degree of incompleteness, and so even if a genuine counterpart was observed it may have been discarded because the host lacks identification.
Either case implies that faint flares or flares with faint hosts are not included in the current counterpart candidate sample.
The respective effects on the inferred parameters is similarly obscured at this time, as it is unknown whether increasing completeness would recover a proportionally larger amount of GW counterpart or background flares.
At any rate, at this point we can only claim that less than a few percent of LVK BBHs produce the bright flares observed in the existing AGN catalogs, which are biased towards the brightest objects, and that deeper follow-up and catalogs are required to extend the limits of these kinds of analyses.
This line of thought has some synergy with existing theory, as multiple teams have predicted that BBH mergers and counterpart observability are hampered as central BH mass increases \citep{mckernan_ram-pressure_2019,rodriguez-ramirez_optical_2024,delfavero_mcfacts_2024}. Note that \citet{Rowan:2024lla} predict that the majority of BBH mergers in AGN disks should occur around SMBHs of $\sim 10^7~M_\odot$, a sweet spot between the rare, high-mass AGNs and the small numbers of BHs around dwarf AGNs. The flaring AGNs in \citet{graham_light_2023} are all $\gtrsim 10^8~M_\odot$, which is reasonable since most AGNs in Milliquas are biased towards the most luminous and massive AGNs. It is therefore possible that the AGNs targeted as part of this search were not associated to the GW events in question, but this does not imply that the BBH mergers did not originate in AGN disks nor that there was no EM counterpart produced. These findings highlight the need to produce AGN catalogs that include objects down to masses of $\lesssim 10^7~M_\odot$, and extend GW follow up campaigns to include such dimmer hosts.

To enable the best use of observing resources, more effort must be dedicated toward elevating the rate of counterparts per follow-up by making proficient use of the GW event data.
Current theories predict that flare energies and timescales depend on the mass of the remnant BH (e.g. \citealt{rodriguez-ramirez_optical_2024}), but the nature of the relationship varies for different flaring mechanisms.
Such a miasma of possibilities is difficult to overcome, but a clearer path forward is possible if flare models are developed in particular consideration of GW observables such as chirp mass and spin.
Of course, such tools would be particularly useful if low-latency spin estimations were made available in the initial GW alert stream along with chirp masses, and so progress in this direction must be made in concert to enable the community to perform the best science possible.
Advancing along these two joined paths will greatly reduce the cost of discovering the first BBH counterpart, or of associatively gathering enough evidence that such a possibility can be excluded.

We emphasize the utility of survey and archival counterpart searches such as that of \citet{graham_light_2023} as important probes of multimessenger BBHs to improve understanding of the uncertain landscape of these phenomena.
The number of GW-observed BBHs roughly tripled during O4, and so one can expect that an enlarged sample of coincident AGN flares will be discernible from contemporary data.
Such a composite dataset will be useful in analyses such as these to further constrain the nature of BBH counterparts, which will help enable the dedicated follow-up mentioned above.
Looking toward the future, the overlap of the Rubin Observatory Legacy Survey of Space and Time (LSST) and LVK O5 will provide a wealth of new information, with LSST probing several magnitudes deeper than ZTF through a dedicated program for BBH mergers follow-up \citep{RubinToO}, and with O5 predicted to detect mergers at roughly ten times the rate of O4.
Such large datasets will be all the more suited for statistical analyses relying on a representative sample of events, and will enable additional science.
In particular, the method of \citet{palmese_ligovirgo_2021} employed here has also been shown to probe cosmological parameters such as the Hubble constant $H_0$ through a statistical standard siren method; while the present work with its uncertain associations between GWs and AGN flares does not possess sufficient information for significant constraints, the breadth of future datasets are expected to alleviate this deficiency \citep{bom_standard_2024} and allow steps towards resolving open questions in the field.

\begin{acknowledgements}

We thank Colin Burke, Saavik Ford, Barry McKernan, and Matthew Graham for useful discussion. TC and AP acknowledge that this material is based upon work supported by NSF Grant No. 2308193. This work used resources on the Vera Cluster at the Pittsburgh Supercomputing Center (PSC).
We thank the PSC staff for help with setting up our software on the Vera Cluster.
We thank our anonymous review for their comments which led to significant improvement of this work.

This research has made use of data or software obtained from the Gravitational Wave Open Science Center (gwosc.org), a service of the LIGO Scientific Collaboration, the Virgo Collaboration, and KAGRA.
This material is based upon work supported by NSF's LIGO Laboratory which is a major facility fully funded by the National Science Foundation, as well as the Science and Technology Facilities Council (STFC) of the United Kingdom, the Max-Planck-Society (MPS), and the State of Niedersachsen/Germany for support of the construction of Advanced LIGO and construction and operation of the GEO600 detector.
Additional support for Advanced LIGO was provided by the Australian Research Council. Virgo is funded, through the European Gravitational Observatory (EGO), by the French Centre National de Recherche Scientifique (CNRS), the Italian Istituto Nazionale di Fisica Nucleare (INFN) and the Dutch Nikhef, with contributions by institutions from Belgium, Germany, Greece, Hungary, Ireland, Japan, Monaco, Poland, Portugal, Spain. KAGRA is supported by Ministry of Education, Culture, Sports, Science and Technology (MEXT), Japan Society for the Promotion of Science (JSPS) in Japan; National Research Foundation (NRF) and Ministry of Science and ICT (MSIT) in Korea; Academia Sinica (AS) and National Science and Technology Council (NSTC) in Taiwan.

This research has made use of the NASA/IPAC Infrared Science Archive, which is funded by the National Aeronautics and Space Administration and operated by the California Institute of Technology.

Based on observations obtained with the Samuel Oschin Telescope 48-inch and the 60-inch Telescope at the Palomar Observatory as part of the Zwicky Transient Facility project. ZTF is supported by the National Science Foundation under Grants No. AST-1440341 and AST-2034437 and a collaboration including current partners Caltech, IPAC, the Oskar Klein Center at Stockholm University, the University of Maryland, University of California, Berkeley , the University of Wisconsin at Milwaukee, University of Warwick, Ruhr University, Cornell University, Northwestern University and Drexel University. Operations are conducted by COO, IPAC, and UW.

\software{
astropy \citep{astropycollaborationAstropyCommunityPython2013, astropycollaborationAstropyProjectBuilding2018, astropycollaborationAstropyProjectSustaining2022},
emcee \citep{foreman-mackey_emcee_2013},
healpy \citep{gorskiHEALPixFrameworkHighResolution2005, zoncaHealpyEqualArea2019},
ligo.skymap \citep{singerGoingDistanceMapping2016},
matplotlib \citep{hunterMatplotlib2DGraphics2007},
numpy \citep{harrisArrayProgrammingNumPy2020},
pandas \citep{mckinneyDataStructuresStatistical2010}
}

\end{acknowledgements}

\facility{IRSA}

\bibliography{lambda_bbhagn}
\bibliographystyle{aasjournal}


\startlongtable
\begin{deluxetable*}{ccccccc}
    \label{tab:gws}
    \tablecaption{
        Parameters for the 76 gravitational wave events used in this study, reproduced from \citealt{abbott_gwtc-2_2021, abbott_gwtc-3_2023}.
        For all events, the 90\% areas were calculated with the \texttt{ligo.skymap.postprocess.crossmatch.crossmatch} function, using IMRPhenomXPHM waveform skymaps.
    }
    \tablehead{
        \colhead{Event ID} & \colhead{90\% Area} & \colhead{$d_L$} & \colhead{$m_1$} & \colhead{$m_2$} & \colhead{$m_{\rm fin}$} & \colhead{$\chi_{\rm eff}$} \\
        & deg$^2$ & Mpc & $M_\odot$ & $M_\odot$ & $M_\odot$ & 
    }
    \startdata
		GW190403\_051519 & 2731 & $8280_{-4290}^{+6720}$ & $85.0_{-33.0}^{+27.8}$ & $20.0_{-8.4}^{+26.3}$ & $102.2_{-24.3}^{+26.3}$ & $0.68_{-0.43}^{+0.16}$ \\ 
		GW190408\_181802 & 271 & $1540_{-620}^{+440}$ & $24.8_{-3.5}^{+5.4}$ & $18.5_{-4.0}^{+3.3}$ & $41.4_{-2.9}^{+3.9}$ & $-0.03_{-0.17}^{+0.13}$ \\ 
		GW190412 & 25 & $720_{-220}^{+240}$ & $27.7_{-6.0}^{+6.0}$ & $9.0_{-1.4}^{+2.0}$ & $35.6_{-4.5}^{+4.8}$ & $0.21_{-0.13}^{+0.12}$ \\ 
		GW190413\_052954 & 668 & $3320_{-1400}^{+1910}$ & $33.7_{-6.4}^{+10.4}$ & $24.2_{-7.0}^{+6.5}$ & $55.5_{-7.3}^{+10.1}$ & $-0.04_{-0.32}^{+0.27}$ \\ 
		GW190413\_134308 & 562 & $3800_{-1830}^{+2480}$ & $51.3_{-12.6}^{+16.6}$ & $30.4_{-12.7}^{+11.7}$ & $78.0_{-11.5}^{+16.1}$ & $-0.01_{-0.38}^{+0.28}$ \\ 
		GW190421\_213856 & 1237 & $2590_{-1240}^{+1490}$ & $42.0_{-7.4}^{+10.1}$ & $32.0_{-9.8}^{+8.3}$ & $70.5_{-9.0}^{+12.4}$ & $-0.1_{-0.27}^{+0.21}$ \\ 
		GW190426\_190642 & 4559 & $4580_{-2280}^{+3400}$ & $105.5_{-24.1}^{+45.3}$ & $76.0_{-36.5}^{+26.2}$ & $172.9_{-33.6}^{+37.7}$ & $0.23_{-0.41}^{+0.42}$ \\ 
		GW190503\_185404 & 103 & $1520_{-600}^{+630}$ & $41.3_{-7.7}^{+10.3}$ & $28.3_{-9.2}^{+7.5}$ & $66.5_{-7.9}^{+9.4}$ & $-0.05_{-0.3}^{+0.23}$ \\ 
		GW190512\_180714 & 274 & $1460_{-590}^{+510}$ & $23.2_{-5.6}^{+5.6}$ & $12.5_{-2.6}^{+3.5}$ & $34.3_{-3.4}^{+4.1}$ & $0.02_{-0.14}^{+0.13}$ \\ 
		GW190513\_205428 & 448 & $2210_{-810}^{+990}$ & $36.0_{-9.7}^{+10.6}$ & $18.3_{-4.7}^{+7.4}$ & $52.1_{-6.6}^{+8.8}$ & $0.16_{-0.22}^{+0.29}$ \\ 
		GW190514\_065416 & 3186 & $3890_{-2070}^{+2610}$ & $40.9_{-9.3}^{+17.3}$ & $28.4_{-10.1}^{+10.0}$ & $66.4_{-11.5}^{+19.0}$ & $-0.08_{-0.35}^{+0.29}$ \\ 
		GW190517\_055101 & 365 & $1790_{-880}^{+1750}$ & $39.2_{-9.2}^{+13.9}$ & $24.0_{-7.9}^{+7.4}$ & $60.1_{-9.4}^{+9.9}$ & $0.49_{-0.28}^{+0.21}$ \\ 
		GW190519\_153544 & 672 & $2600_{-960}^{+1720}$ & $65.1_{-11.0}^{+10.8}$ & $40.8_{-12.7}^{+11.5}$ & $100.0_{-12.9}^{+13.0}$ & $0.33_{-0.24}^{+0.2}$ \\ 
		GW190521 & 1021 & $3310_{-1800}^{+2790}$ & $98.4_{-21.7}^{+33.6}$ & $57.2_{-30.1}^{+27.1}$ & $147.4_{-16.0}^{+40.0}$ & $-0.14_{-0.45}^{+0.5}$ \\ 
		GW190521\_074359 & 469 & $1080_{-530}^{+580}$ & $43.4_{-5.5}^{+5.8}$ & $33.4_{-6.8}^{+5.2}$ & $72.6_{-5.4}^{+6.5}$ & $0.1_{-0.13}^{+0.13}$ \\ 
		GW190527\_092055 & 3640 & $2520_{-1230}^{+2080}$ & $35.6_{-8.0}^{+18.7}$ & $22.2_{-8.7}^{+9.0}$ & $55.5_{-8.5}^{+17.9}$ & $0.1_{-0.22}^{+0.22}$ \\ 
		GW190602\_175927 & 739 & $2840_{-1280}^{+1930}$ & $71.8_{-14.6}^{+18.1}$ & $44.8_{-19.6}^{+15.5}$ & $110.5_{-13.9}^{+17.9}$ & $0.12_{-0.28}^{+0.25}$ \\ 
		GW190620\_030421 & 6443 & $2910_{-1320}^{+1710}$ & $58.0_{-13.3}^{+19.2}$ & $35.0_{-14.5}^{+13.1}$ & $88.0_{-12.4}^{+17.2}$ & $0.34_{-0.29}^{+0.22}$ \\ 
		GW190630\_185205 & 960 & $870_{-360}^{+530}$ & $35.1_{-5.5}^{+6.5}$ & $24.0_{-5.2}^{+5.5}$ & $56.6_{-4.5}^{+4.4}$ & $0.1_{-0.13}^{+0.14}$ \\ 
		GW190701\_203306 & 43 & $2090_{-740}^{+770}$ & $54.1_{-8.0}^{+12.6}$ & $40.5_{-12.1}^{+8.7}$ & $90.2_{-8.9}^{+11.2}$ & $-0.08_{-0.31}^{+0.23}$ \\ 
		GW190706\_222641 & 2596 & $3630_{-2000}^{+2600}$ & $74.0_{-16.9}^{+20.1}$ & $39.4_{-15.4}^{+18.4}$ & $107.3_{-15.9}^{+25.2}$ & $0.28_{-0.31}^{+0.25}$ \\ 
		GW190707\_093326 & 893 & $850_{-400}^{+340}$ & $12.1_{-2.0}^{+2.6}$ & $7.9_{-1.3}^{+1.6}$ & $19.2_{-1.2}^{+1.7}$ & $-0.04_{-0.09}^{+0.1}$ \\ 
		GW190708\_232457 & 11032 & $930_{-390}^{+310}$ & $19.8_{-4.3}^{+4.3}$ & $11.6_{-2.0}^{+3.1}$ & $30.1_{-2.1}^{+2.9}$ & $0.05_{-0.1}^{+0.1}$ \\ 
		GW190719\_215514 & 3564 & $3730_{-2070}^{+3120}$ & $36.6_{-11.1}^{+42.1}$ & $19.9_{-9.3}^{+10.0}$ & $54.5_{-11.1}^{+38.3}$ & $0.25_{-0.32}^{+0.33}$ \\ 
		GW190720\_000836 & 35 & $770_{-260}^{+650}$ & $14.2_{-3.3}^{+5.6}$ & $7.5_{-1.8}^{+2.2}$ & $20.8_{-2.0}^{+3.9}$ & $0.19_{-0.11}^{+0.14}$ \\ 
		GW190725\_174728 & 2142 & $1030_{-430}^{+520}$ & $11.8_{-3.0}^{+10.1}$ & $6.3_{-2.5}^{+2.1}$ & $17.6_{-1.8}^{+7.7}$ & $-0.04_{-0.16}^{+0.36}$ \\ 
		GW190727\_060333 & 100 & $3070_{-1230}^{+1300}$ & $38.9_{-6.0}^{+8.9}$ & $30.2_{-8.3}^{+6.5}$ & $65.4_{-7.3}^{+9.5}$ & $0.09_{-0.27}^{+0.25}$ \\ 
		GW190728\_064510 & 321 & $880_{-380}^{+260}$ & $12.5_{-2.3}^{+6.9}$ & $8.0_{-2.6}^{+1.7}$ & $19.7_{-1.4}^{+4.4}$ & $0.13_{-0.07}^{+0.19}$ \\ 
		GW190731\_140936 & 3532 & $3330_{-1770}^{+2350}$ & $41.8_{-9.1}^{+12.7}$ & $29.0_{-9.9}^{+10.2}$ & $67.4_{-10.8}^{+15.3}$ & $0.07_{-0.25}^{+0.28}$ \\ 
		GW190803\_022701 & 1012 & $3190_{-1470}^{+1630}$ & $37.7_{-6.7}^{+9.8}$ & $27.6_{-8.5}^{+7.6}$ & $62.1_{-7.6}^{+11.2}$ & $-0.01_{-0.28}^{+0.23}$ \\ 
		GW190805\_211137 & 1538 & $6130_{-3080}^{+3720}$ & $46.2_{-11.2}^{+15.4}$ & $30.6_{-11.3}^{+11.8}$ & $72.4_{-13.2}^{+18.2}$ & $0.37_{-0.39}^{+0.29}$ \\ 
		GW190814 & 22 & $230_{-50}^{+40}$ & $23.3_{-1.4}^{+1.4}$ & $2.6_{-0.1}^{+0.1}$ & $25.7_{-1.3}^{+1.3}$ & $0.0_{-0.07}^{+0.07}$ \\ 
		GW190828\_063405 & 340 & $2070_{-920}^{+650}$ & $31.9_{-4.1}^{+5.4}$ & $25.8_{-5.3}^{+4.9}$ & $54.3_{-4.0}^{+7.3}$ & $0.15_{-0.16}^{+0.15}$ \\ 
		GW190828\_065509 & 593 & $1540_{-650}^{+690}$ & $23.7_{-6.7}^{+6.8}$ & $10.4_{-2.2}^{+3.8}$ & $33.0_{-4.3}^{+5.3}$ & $0.05_{-0.17}^{+0.16}$ \\ 
		GW190910\_112807 & 8305 & $1520_{-630}^{+1090}$ & $43.8_{-6.8}^{+7.6}$ & $34.2_{-7.3}^{+6.6}$ & $74.4_{-8.6}^{+8.5}$ & $0.0_{-0.2}^{+0.17}$ \\ 
		GW190915\_235702 & 432 & $1750_{-650}^{+710}$ & $32.6_{-4.9}^{+8.8}$ & $24.5_{-5.8}^{+4.9}$ & $54.7_{-5.0}^{+6.6}$ & $-0.03_{-0.24}^{+0.19}$ \\ 
		GW190916\_200658 & 2368 & $4940_{-2380}^{+3710}$ & $43.8_{-12.6}^{+19.9}$ & $23.3_{-10.0}^{+12.5}$ & $65.0_{-12.6}^{+17.3}$ & $0.2_{-0.31}^{+0.33}$ \\ 
		GW190917\_114630 & 1687 & $720_{-310}^{+300}$ & $9.7_{-3.9}^{+3.4}$ & $2.1_{-0.4}^{+1.1}$ & $11.6_{-2.9}^{+3.1}$ & $-0.08_{-0.43}^{+0.21}$ \\ 
		GW190924\_021846 & 376 & $550_{-220}^{+220}$ & $8.8_{-1.8}^{+4.3}$ & $5.1_{-1.5}^{+1.2}$ & $13.3_{-0.9}^{+3.0}$ & $0.03_{-0.08}^{+0.2}$ \\ 
		GW190925\_232845 & 876 & $930_{-350}^{+460}$ & $20.8_{-2.9}^{+6.5}$ & $15.5_{-3.6}^{+2.5}$ & $34.9_{-2.6}^{+3.5}$ & $0.09_{-0.15}^{+0.16}$ \\ 
		GW190926\_050336 & 2015 & $3280_{-1730}^{+3400}$ & $41.1_{-12.5}^{+20.8}$ & $20.4_{-8.2}^{+11.4}$ & $59.6_{-11.8}^{+22.1}$ & $-0.02_{-0.32}^{+0.25}$ \\ 
		GW190929\_012149 & 1651 & $3130_{-1370}^{+2510}$ & $66.3_{-16.6}^{+21.6}$ & $26.8_{-10.6}^{+14.7}$ & $90.3_{-14.6}^{+22.3}$ & $-0.03_{-0.28}^{+0.23}$ \\ 
		GW190930\_133541 & 1493 & $770_{-320}^{+320}$ & $14.2_{-4.0}^{+8.0}$ & $6.9_{-2.1}^{+2.4}$ & $20.2_{-2.0}^{+6.1}$ & $0.19_{-0.16}^{+0.22}$ \\ 
		GW191103\_012549 & 2171 & $990_{-470}^{+500}$ & $11.8_{-2.2}^{+6.2}$ & $7.9_{-2.4}^{+1.7}$ & $19.0_{-1.7}^{+3.8}$ & $0.21_{-0.1}^{+0.16}$ \\ 
		GW191105\_143521 & 641 & $1150_{-480}^{+430}$ & $10.7_{-1.6}^{+3.7}$ & $7.7_{-1.9}^{+1.4}$ & $17.6_{-1.2}^{+2.1}$ & $-0.02_{-0.09}^{+0.13}$ \\ 
		GW191109\_010717 & 1649 & $1290_{-650}^{+1130}$ & $65.0_{-11.0}^{+11.0}$ & $47.0_{-13.0}^{+15.0}$ & $107.0_{-15.0}^{+18.0}$ & $-0.29_{-0.31}^{+0.42}$ \\ 
		GW191113\_071753 & 2484 & $1370_{-620}^{+1150}$ & $29.0_{-14.0}^{+12.0}$ & $5.9_{-1.3}^{+4.4}$ & $34.0_{-10.0}^{+11.0}$ & $0.0_{-0.29}^{+0.37}$ \\ 
		GW191126\_115259 & 1378 & $1620_{-740}^{+740}$ & $12.1_{-2.2}^{+5.5}$ & $8.3_{-2.4}^{+1.9}$ & $19.6_{-2.0}^{+3.5}$ & $0.21_{-0.11}^{+0.15}$ \\ 
		GW191127\_050227 & 983 & $3400_{-1900}^{+3100}$ & $53.0_{-20.0}^{+47.0}$ & $24.0_{-14.0}^{+17.0}$ & $76.0_{-21.0}^{+39.0}$ & $0.18_{-0.36}^{+0.34}$ \\ 
		GW191129\_134029 & 856 & $790_{-330}^{+260}$ & $10.7_{-2.1}^{+4.1}$ & $6.7_{-1.7}^{+1.5}$ & $16.8_{-1.2}^{+2.5}$ & $0.06_{-0.08}^{+0.16}$ \\ 
		GW191204\_110529 & 3675 & $1900_{-1100}^{+1700}$ & $27.3_{-5.9}^{+10.8}$ & $19.2_{-6.0}^{+5.5}$ & $45.0_{-7.5}^{+8.7}$ & $0.05_{-0.26}^{+0.25}$ \\ 
		GW191204\_171526 & 256 & $640_{-260}^{+200}$ & $11.7_{-1.7}^{+3.3}$ & $8.4_{-1.7}^{+1.3}$ & $19.18_{-0.93}^{+1.71}$ & $0.16_{-0.05}^{+0.08}$ \\ 
		GW191215\_223052 & 586 & $1930_{-860}^{+890}$ & $24.9_{-4.1}^{+7.1}$ & $18.1_{-4.1}^{+3.8}$ & $41.4_{-4.1}^{+5.1}$ & $-0.04_{-0.21}^{+0.17}$ \\ 
		GW191216\_213338 & 206 & $340_{-130}^{+120}$ & $12.1_{-2.2}^{+4.6}$ & $7.7_{-1.9}^{+1.6}$ & $18.87_{-0.93}^{+2.81}$ & $0.11_{-0.06}^{+0.13}$ \\ 
		GW191222\_033537 & 2168 & $3000_{-1700}^{+1700}$ & $45.1_{-8.0}^{+10.9}$ & $34.7_{-10.5}^{+9.3}$ & $75.5_{-9.9}^{+15.3}$ & $-0.04_{-0.25}^{+0.2}$ \\ 
		GW191230\_180458 & 1086 & $4300_{-1900}^{+2100}$ & $49.4_{-9.6}^{+14.0}$ & $37.0_{-12.0}^{+11.0}$ & $82.0_{-11.0}^{+17.0}$ & $-0.05_{-0.31}^{+0.26}$ \\ 
		GW200112\_155838 & 3200 & $1250_{-460}^{+430}$ & $35.6_{-4.5}^{+6.7}$ & $28.3_{-5.9}^{+4.4}$ & $60.8_{-4.3}^{+5.3}$ & $0.06_{-0.15}^{+0.15}$ \\ 
		GW200128\_022011 & 2415 & $3400_{-1800}^{+2100}$ & $42.2_{-8.1}^{+11.6}$ & $32.6_{-9.2}^{+9.5}$ & $71.0_{-11.0}^{+16.0}$ & $0.12_{-0.25}^{+0.24}$ \\ 
		GW200129\_065458 & 31 & $890_{-370}^{+260}$ & $34.5_{-3.1}^{+9.9}$ & $29.0_{-9.3}^{+3.3}$ & $60.2_{-3.2}^{+4.1}$ & $0.11_{-0.16}^{+0.11}$ \\ 
		GW200202\_154313 & 150 & $410_{-160}^{+150}$ & $10.1_{-1.4}^{+3.5}$ & $7.3_{-1.7}^{+1.1}$ & $16.76_{-0.66}^{+1.87}$ & $0.04_{-0.06}^{+0.13}$ \\ 
		GW200208\_130117 & 30 & $2230_{-850}^{+1020}$ & $37.7_{-6.2}^{+9.3}$ & $27.4_{-7.3}^{+6.3}$ & $62.5_{-6.4}^{+7.5}$ & $-0.07_{-0.27}^{+0.21}$ \\ 
		GW200208\_222617 & 2040 & $4100_{-2000}^{+4400}$ & $51.0_{-30.0}^{+103.0}$ & $12.3_{-5.5}^{+9.2}$ & $61.0_{-26.0}^{+99.0}$ & $0.45_{-0.46}^{+0.42}$ \\ 
		GW200209\_085452 & 877 & $3400_{-1800}^{+1900}$ & $35.6_{-6.8}^{+10.5}$ & $27.1_{-7.8}^{+7.8}$ & $59.9_{-8.9}^{+13.1}$ & $-0.12_{-0.3}^{+0.24}$ \\ 
		GW200210\_092254 & 1387 & $940_{-340}^{+430}$ & $24.1_{-4.6}^{+7.5}$ & $2.83_{-0.42}^{+0.47}$ & $26.7_{-4.3}^{+7.2}$ & $0.02_{-0.21}^{+0.22}$ \\ 
		GW200216\_220804 & 2924 & $3800_{-2000}^{+3000}$ & $51.0_{-13.0}^{+22.0}$ & $30.0_{-16.0}^{+14.0}$ & $78.0_{-13.0}^{+19.0}$ & $0.1_{-0.36}^{+0.34}$ \\ 
		GW200219\_094415 & 781 & $3400_{-1500}^{+1700}$ & $37.5_{-6.9}^{+10.1}$ & $27.9_{-8.4}^{+7.4}$ & $62.2_{-7.8}^{+11.7}$ & $-0.08_{-0.29}^{+0.23}$ \\ 
		GW200220\_061928 & 4477 & $6000_{-3100}^{+4800}$ & $87.0_{-23.0}^{+40.0}$ & $61.0_{-25.0}^{+26.0}$ & $141.0_{-31.0}^{+51.0}$ & $0.06_{-0.38}^{+0.4}$ \\ 
		GW200220\_124850 & 3129 & $4000_{-2200}^{+2800}$ & $38.9_{-8.6}^{+14.1}$ & $27.9_{-9.0}^{+9.2}$ & $64.0_{-11.0}^{+16.0}$ & $-0.07_{-0.33}^{+0.27}$ \\ 
		GW200224\_222234 & 42 & $1710_{-650}^{+500}$ & $40.0_{-4.5}^{+6.7}$ & $32.7_{-7.2}^{+4.8}$ & $68.7_{-4.8}^{+6.7}$ & $0.1_{-0.16}^{+0.15}$ \\ 
		GW200225\_060421 & 498 & $1150_{-530}^{+510}$ & $19.3_{-3.0}^{+5.0}$ & $14.0_{-3.5}^{+2.8}$ & $32.1_{-2.8}^{+3.5}$ & $-0.12_{-0.28}^{+0.17}$ \\ 
		GW200302\_015811 & 6016 & $1480_{-700}^{+1020}$ & $37.8_{-8.5}^{+8.7}$ & $20.0_{-5.7}^{+8.1}$ & $55.5_{-6.6}^{+8.9}$ & $0.01_{-0.26}^{+0.25}$ \\ 
		GW200306\_093714 & 3907 & $2100_{-1100}^{+1700}$ & $28.3_{-7.7}^{+17.1}$ & $14.8_{-6.4}^{+6.5}$ & $41.7_{-6.9}^{+12.3}$ & $0.32_{-0.46}^{+0.28}$ \\ 
		GW200308\_173609 & 25292 & $7100_{-4400}^{+13900}$ & $60.0_{-29.0}^{+166.0}$ & $24.0_{-13.0}^{+36.0}$ & $88.0_{-47.0}^{+169.0}$ & $0.16_{-0.49}^{+0.58}$ \\ 
		GW200311\_115853 & 35 & $1170_{-400}^{+280}$ & $34.2_{-3.8}^{+6.4}$ & $27.7_{-5.9}^{+4.1}$ & $59.0_{-3.9}^{+4.8}$ & $-0.02_{-0.2}^{+0.16}$ \\ 
		GW200316\_215756 & 187 & $1120_{-440}^{+480}$ & $13.1_{-2.9}^{+10.2}$ & $7.8_{-2.9}^{+2.0}$ & $20.2_{-1.9}^{+7.4}$ & $0.13_{-0.1}^{+0.27}$ \\ 
		GW200322\_091133 & 28704 & $3500_{-2200}^{+12500}$ & $38.0_{-22.0}^{+130.0}$ & $11.3_{-6.0}^{+24.3}$ & $48.0_{-22.0}^{+132.0}$ & $0.27_{-0.58}^{+0.54}$ \\ 
\enddata
\end{deluxetable*}

\begin{deluxetable}{ccccc}
    \label{tab:flares}
    \tablecaption{
        Parameters for the 17 AGN flares used in this study.
        Host redshifts are reproduced from \citealt{veronesi_agn-flares_2025}.
        Onset dates ${\rm MJD}_0$ are calculated as ${\rm MJD}_{\rm peak} - 3 \sigma_{\rm rise}$, where the two parameters are taken from fitting a Gaussian rise-exponential decay model to the ZTF $g$-band lightcurve \citep{graham_light_2023}: ${\rm MJD}_{\rm peak}$ is the fit time of flare maximum, and $\sigma_{\rm rise}$ is the fit standard deviation of the Gaussian rise.
        Note that the RA and declination of each flare are recoverable from the respective Event ID, which is the SDSS designation of the host galaxy.
    }
    \tablehead{
        \colhead{Event ID} & \colhead{Redshift} & \colhead{${\rm MJD}_{\rm peak}$} & \colhead{$\sigma_{\rm rise}$} & \colhead{${\rm MJD}_0$}
    }
    \startdata
		J120437.98+500024.0 & 0.389 & 58893.6 & 18.4 & 58838.3 \\
		J124942.30+344928.9 & 0.438 & 58662.3 & 4.9 & 58647.7 \\
		J140941.88+552928.1 & 0.074 & 58611.4 & 11.2 & 58577.9 \\
		J143157.51+451544.0 & 0.693 & 58861.1 & 70.7 & 58649.1 \\
		J143536.15+173755.4 & 0.095 & 58676.2 & 7.3 & 58654.4 \\
		J145500.22+321637.1 & 0.177 & 58594.3 & 74.9 & 58369.6 \\
		J152433.35+274311.6 & 0.069 & 58611.6 & 11.1 & 58578.3 \\
		J154342.46+461233.4 & 0.599 & 58966.7 & 32.3 & 58869.8 \\
		J154806.31+291216.3 & 1.090 & 59000.5 & 93.6 & 58719.9 \\
		J160822.16+401217.8 & 0.627 & 58931.0 & 35.1 & 58825.7 \\
		J161833.77+263226.0 & 0.126 & 58596.7 & 6.8 & 58576.2 \\
		J163641.61+092459.2 & 1.155 & 58691.1 & 3.6 & 58680.4 \\
		J181719.95+541910.0 & 0.234 & 58781.4 & 21.9 & 58715.8 \\
		J183412.42+365655.2 & 0.419 & 58694.1 & 15.7 & 58647.0 \\
		J224333.95+760619.2 & 0.353 & 58777.7 & 33.1 & 58678.5 \\
		J233252.05+034559.7 & 1.119 & 58845.0 & 61.1 & 58661.7 \\
		J233746.08-013116.3 & 0.115 & 58701.1 & 1.8 & 58695.6 \\
\enddata
\end{deluxetable}

\end{document}